\RequirePackage{lineno}
\pdfoutput=1
\documentclass[aps,prl,twocolumn,showpacs,showkeys,amsmath,amssymb,superscriptaddress,groupedaddress]{revtex4}  
\usepackage{graphicx}  
\usepackage{subfigure}
\usepackage{dcolumn}   
\usepackage{bm}        
\usepackage{amssymb}   
\usepackage{lineno}
\usepackage{multirow}
\usepackage{indentfirst}
\usepackage{color}     
\usepackage{hyperref}  

\newcommand{\ud}{\mathrm{d}}

\newcommand{\ue}{\mathrm{e}}

\newcommand{\rt}{(7.7\pm3.0)\ \rm{ns}}
\newcommand{\dt}{(36.6\pm2.4)\ \rm{ns}}
\newcommand{\ly}{(1.01\pm0.12)\times10^3\ \rm{photons}/\rm{MeV}}

\begin{document}
\setlength{\parskip}{0\baselineskip}
\title{Separation of Scintillation and Cherenkov Lights in Linear Alkyl Benzene}

\newcommand{\TsingHua}{\affiliation{Department~of~Engineering~Physics, Tsinghua~University, Beijing 100084, China}}
\newcommand{\BNL}{\affiliation{Brookhaven~National~Laboratory, Upton, New York 11973, USA}}

\author{Mohan~Li}\TsingHua
\author{Ziyi~Guo}\TsingHua
\author{Minfang~Yeh}\BNL
\author{Zhe~Wang}\TsingHua
\author{Shaomin~Chen}\TsingHua

\date{November 30, 2015}

\begin{abstract}
To separate scintillation and Cherenkov lights in water-based liquid scintillator detectors is a desired feature for
future neutrino and proton decay researches.
Linear alkyl benzene (LAB) is one important ingredient of a water-based liquid scintillator being developed.
In this paper we observed a good separation of scintillation and Cherenkov lights in an LAB sample.
The rising and decay times of the scintillation light of the LAB were measured to be $\rt$ and $\dt$, respectively,
while the full width [-3$\sigma$, 3$\sigma$] of the Cherenkov light was 12 ns dominated by the time resolution of our photomultiplier tubes. The light yield of the scintillation was measured to be $\ly$.
\end{abstract}

\maketitle

\section{Introduction}
Large-mass, high-resolution, cost-effective detectors will be essential for future neutrino and proton decay experiments~\cite{snowmass}.
Currently we can find two successful options for detector materials: water (heavy water) and liquid scintillator.
Water (heavy-water) Cherenkov detectors have been used by
IMB~\cite{IMB}, Super-Kamiokande~\cite{SK}, SNO~\cite{SNO}, etc.
With this technique the momentum, energy, and direction of a charged particle can be measured, and
muons and electrons can be effectively identified.
But it is limited by the requirement that the momentum of a charged particle must be above its Cherenkov threshold and the low light yield.
Liquid scintillator detectors have been used by KamLAND~\cite{KamLAND}, Borexino~\cite{Borexino}, LSND~\cite{LSND}, Chooz~\cite{DC}, RENO~\cite{RENO}, Daya Bay~\cite{DYB}, etc.
They have much lower detection threshold and higher energy resolution than water Cherenkov detectors, but inefficient in other aspects.

A technique which could combine the different features of both water and liquid scintillator is highly desired.
The concept of water-based liquid scintillator (WbLS) may be found as early as in~\cite{IEEE}.
Recently more efforts for this purpose can be found in~\cite{Minfang, Charac, Damage} and~\cite{Korea}.
With a WbLS, for a charged particle below its Cherenkov threshold,
scintillation light can still be emitted and detected and this will lower the detection threshold of a detector.
Above the threshold,
both Cherenkov and scintillation lights can be detected.
If the excitation and deexcitation responsible for scintillation emission is slower than Cherenkov light emission,
it will be possible to separately measure these two types of lights and probe each particle twice.
The possible improvements with a WbLS for the researches of neutrino physics and proton decay were well recognized in~\cite{ASDC, THEIA}.

With the redundant measurements of a charged particle, a new approach of particle identification can be found. For example, muons and electrons with the same kinetic energy have different amounts of Cherenkov-light emission, which is an intriguing feature for future neutrino and proton decay experiments~\cite{ASDC}.
This feature can be further exploited to separate electrons and gammas for solar neutrino experiments
to suppress the critical external gamma backgrounds~\cite{BX1}.


In this paper we
measured the time profile of scintillation light in a linear alkyl benzene (LAB) sample and
tested the separation of Cherenkov and scintillation lights, since LAB is
one important ingredient of the liquid scintillator of the Daya Bay experiment~\cite{DYB}, etc.,
and also one important ingredient of a WbLS~\cite{Minfang}.
We demonstrated that the separation of Cherenkov and scintillation could be achieved in LAB;
and then such a concept can be further demonstrated by WbLS.
With proper controls of scintillation time and light-yield, the outlook of the separation technique is rather promising for many experiments.

In section~\ref{sec:Detector} we report our apparatus, followed by the data processing in section~\ref{DataProcess}. In section~\ref{sec:Simulation} the simulation tool for our apparatus is described. The measurement results of scintillation light yield and time profile of LAB can be found in section~\ref{sec:Results}.
In the end, we conclude in section~\ref{sec:Conclusion}.

\section{Apparatus}
\label{sec:Detector}
A specific detector was designed to measure the Cherenkov and scintillation lights of vertically incident muons in an LAB sample.

\begin{figure}[h]
\begin{center}
\includegraphics[scale=0.083]{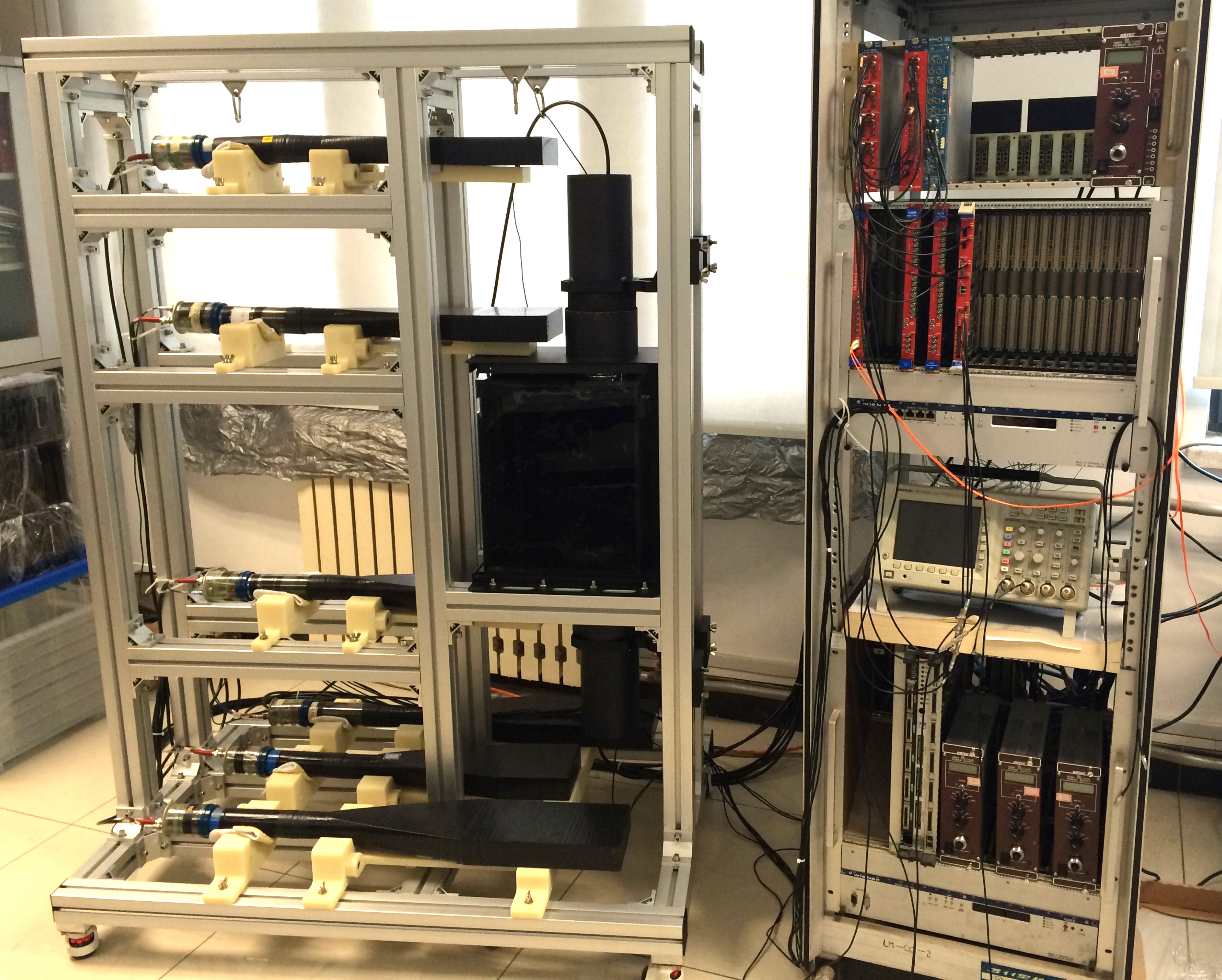}
\end{center}
\caption{Photo of the detector.}
\label{PhotoOfPlatform}
\end{figure}

\begin{figure}[htbp]
\includegraphics[width=8cm]{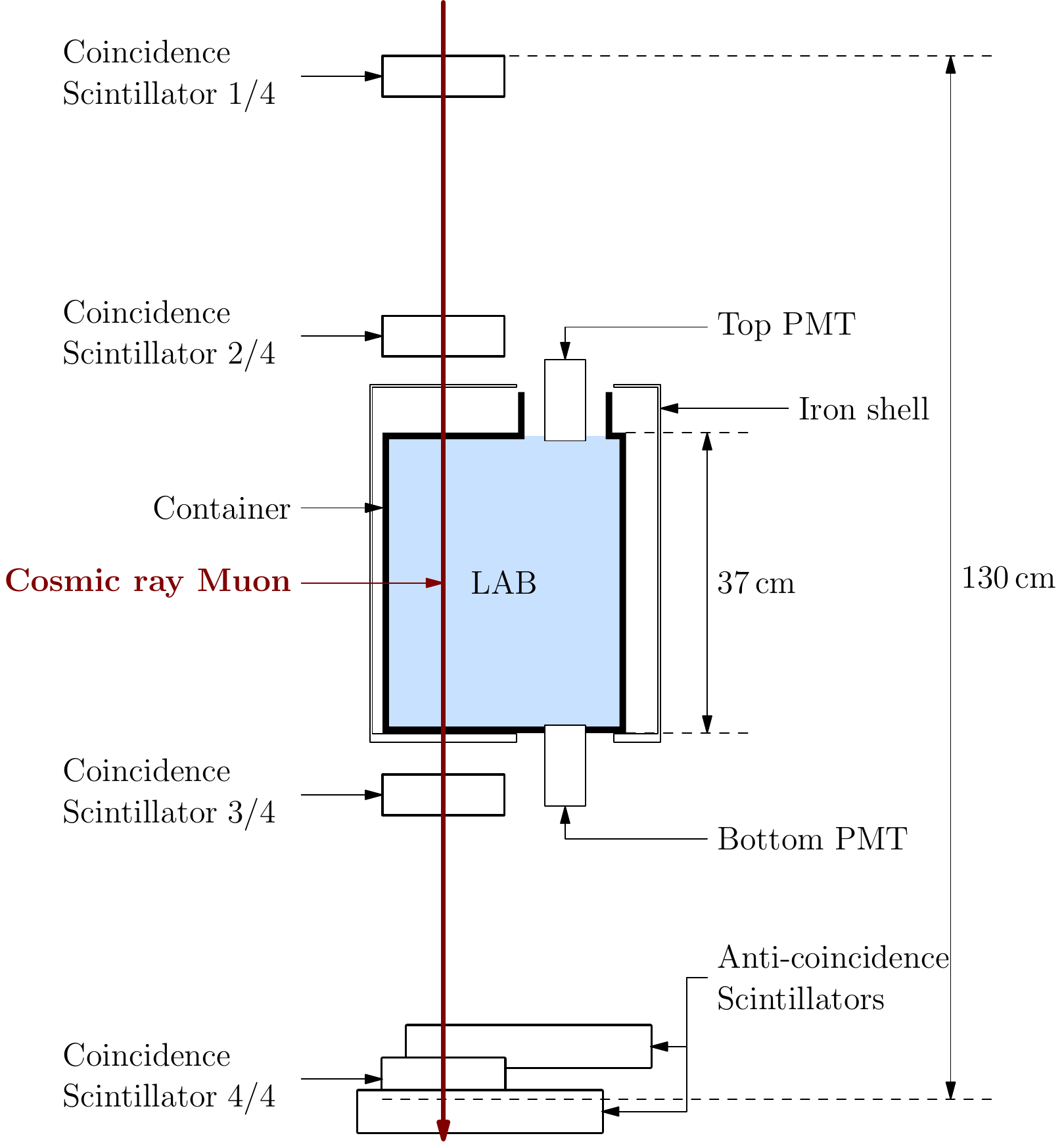}
\caption{Schematic front view of the detector.}
\label{front}
\end{figure}

A photo of the detector is shown in Fig.~\ref{PhotoOfPlatform} and its schematic front view can be found in Fig.~\ref{front}.
Four 15 cm long, 15 cm wide, and 5 cm high plastic scintillators (coincidence scintillators) were placed vertically, whose quadruple-coincidence signals  were used as the triggers of the detector. Another two 30 cm long, 15 cm wide, and 5 cm high plastic scintillators (anti-coincidence scintillators) were placed next to the bottom coincidence scintillator to exclude muon shower events. The light signals of the six plastic scintillators were detected by six photomultiplier tubes (PMT's).

An LAB sample was hold in a 30 cm long, 15 cm wide, and 40 cm high acrylic container having a black and rough inner surface with low reflectivity. The sample was exposed to air and no degassing was performed.
The container was placed between the second and third coincidence scintillators and right on the path of the triggered muons.
The average energy of the muons in our laboratory (sea level) is about 4 GeV~\cite{PDG}, and they will generate both scintillation and Cherenkov lights in LAB.

Two Hamamatsu R1828-01 PMT's were used to measure the scintillation and/or Cherenkov signals.
One PMT was mounted on the top of the container, and the other one was on the bottom.
The photo cathodes of the two PMT's extended into the liquid sample and had a direct contact with the LAB.
For a downward-going muon, the top PMT was expected to measure the scintillation signals only, while the bottom PMT measures the overlapped scintillation and Cherenkov signals.
The typical anode pulse rise time is 1.3 ns according the PMT's data sheet~\cite{Hama}.

After a trigger, an 8 bit 500 MHz flash analog-to-digital converter (FADC, model CAEN VX1721)
opened a 4096 ns window and read out the waveforms of all eight PMT's.
The waveforms were further analyzed offline.

\section{Data Processing}
\label{DataProcess}

\subsection{Signal selection}
Besides the expected signals from single vertically-going muons, two types of background events were recorded:
electronics noise events and muon shower events.
By analyzing the waveforms of the four coincidence channels and the two anti-coincidence channels, the backgrounds can be removed.

\begin{figure}[h]
\begin{center}
\includegraphics[width=9cm]{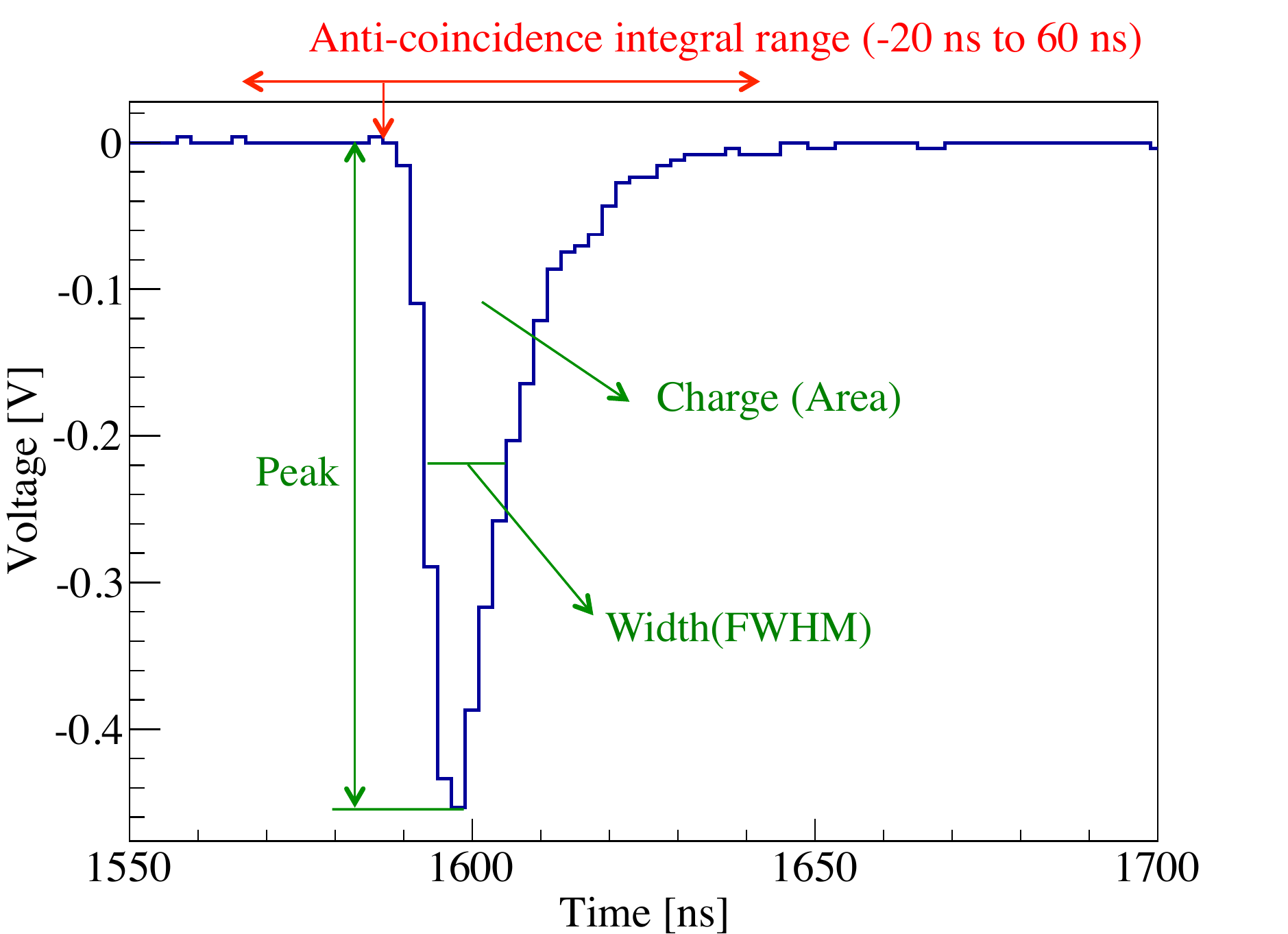}
\end{center}
\caption{A typical waveform of the coincidence channels and three characteristic variables, Peak, Width, and Charge, and
the corresponding anti-coincidence integral window.}
\label{CutParametersDefinition}
\end{figure}

One example waveform of the coincidence channels is shown in Fig.~\ref{CutParametersDefinition}. For each waveform, three characteristic variables, Peak, Width, and Charge (Area), were measured.

Electronics noise events were mainly caused by the coherent noises induced by the power supplies of the PMT's. Their waveforms were much narrower than the real PMT waveforms caused by photons.
Quantities of Peak-to-Charge-ratio and Peak-to-Width-ratio were effective to remove these backgrounds.

Muon shower events should be avoided, because multi tracks may go through our sample container. They were rejected by examining the charge integral in the anti-coincidence channels.
The integral range was [-20 ns, 60 ns] with respect to the rising edge of the coincidence channels as seen in Fig.~\ref{CutParametersDefinition}.
All the events with more significant charges than baseline fluctuations in the anti-coincidence channels were rejected.
The shower backgrounds were further rejected by checking their charges in the coincidence channels.
The energy deposit in each channel should follow a Landau distribution with the assumption of minimum-ionizing particles.
Fig.~\ref{8SignalsInOneEvent} shows the fitting results of them with Landau distributions.
The events with significantly small or large charges than their averages in any of the channels were rejected.

\begin{figure}[h]
\begin{center}
\includegraphics[scale=0.21]{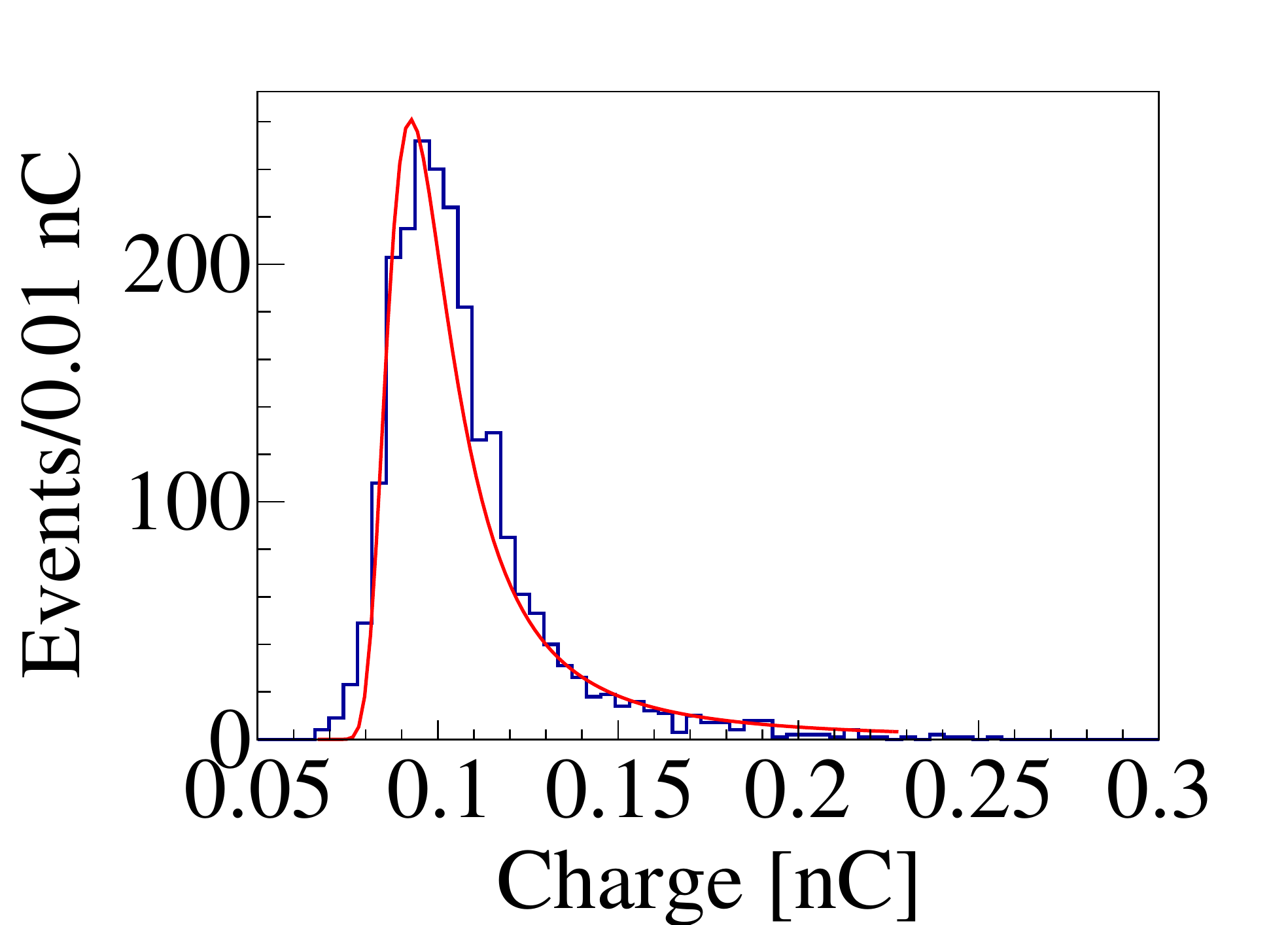}
\includegraphics[scale=0.21]{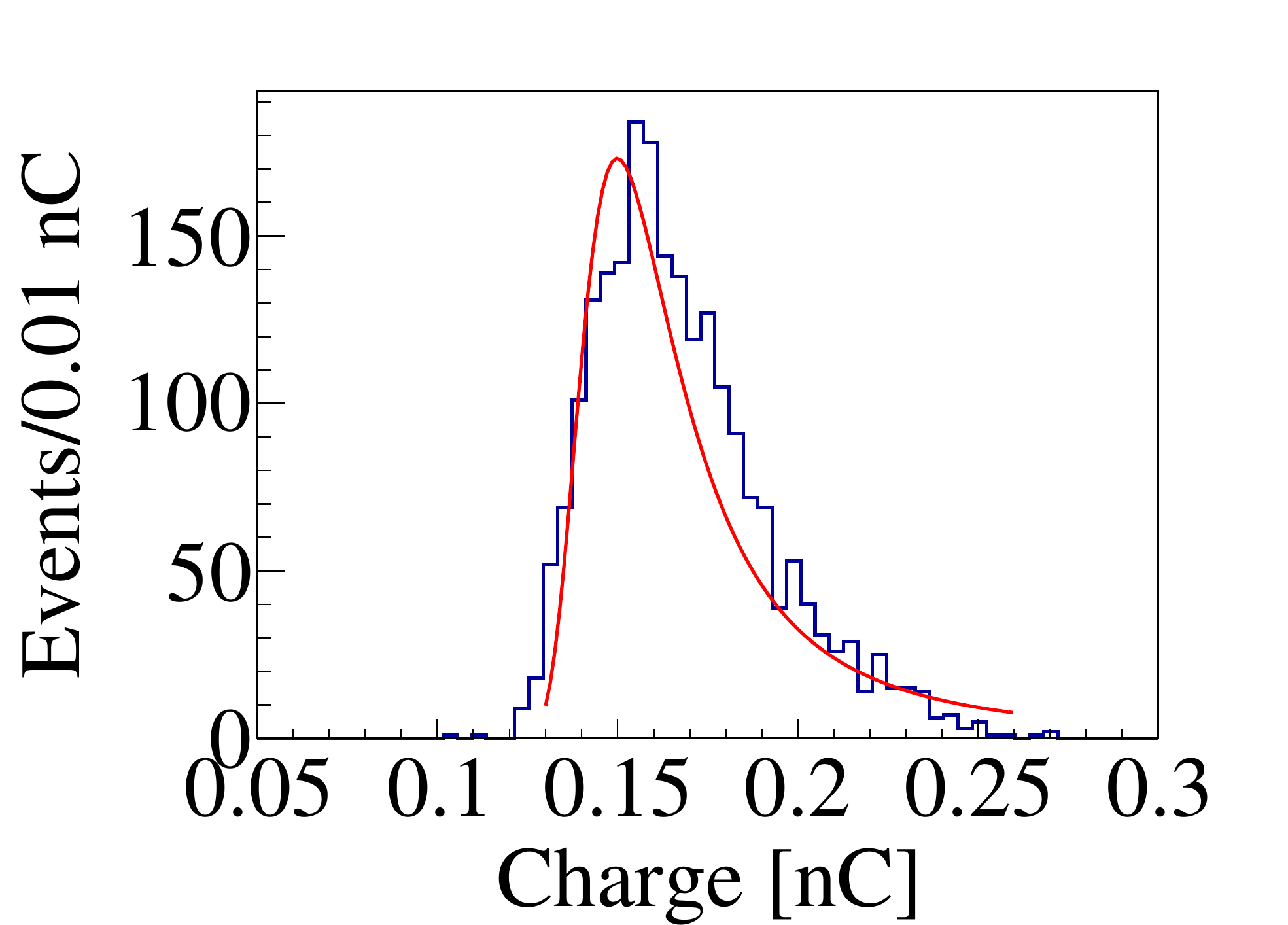}
\includegraphics[scale=0.21]{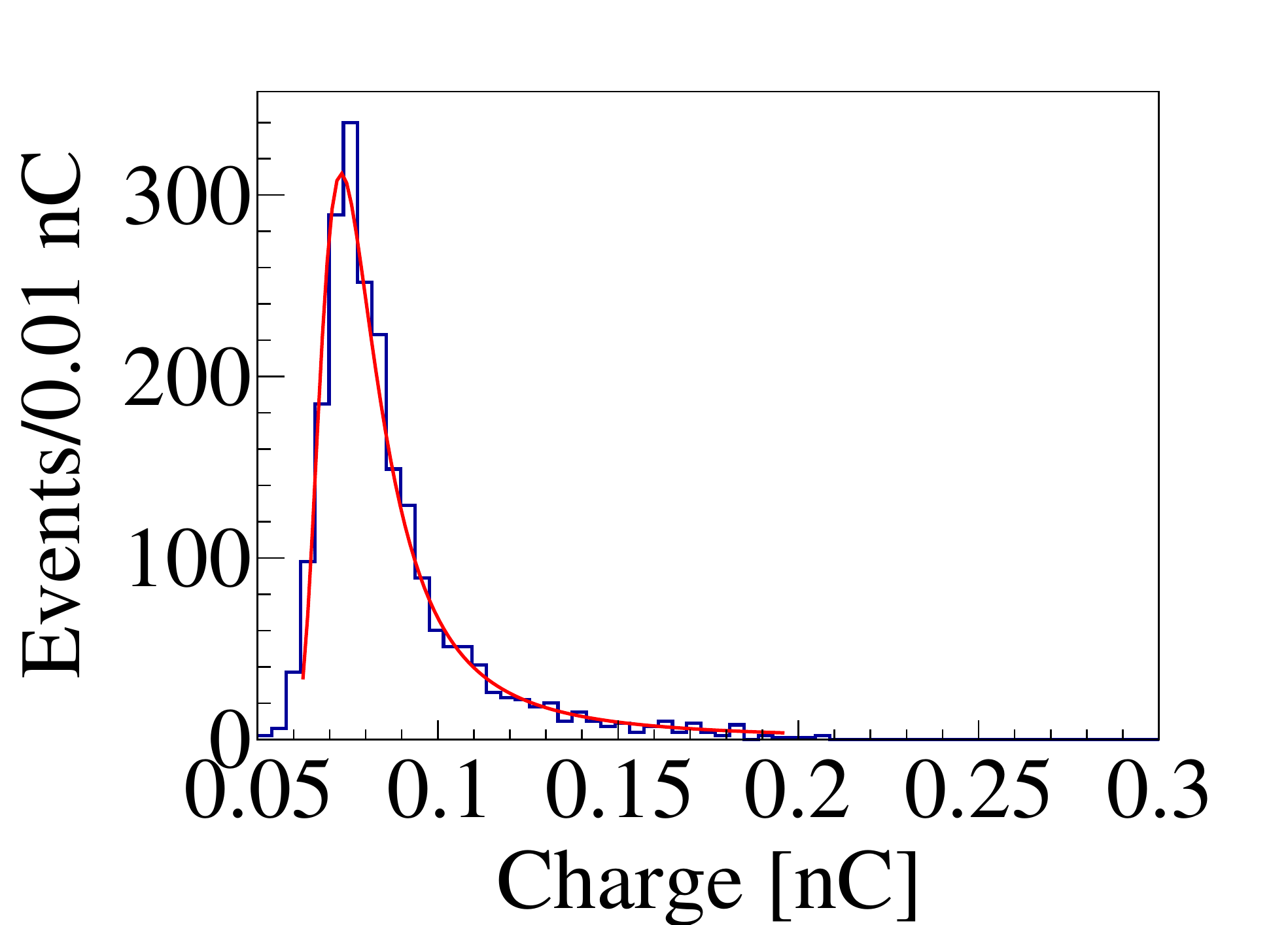}
\includegraphics[scale=0.21]{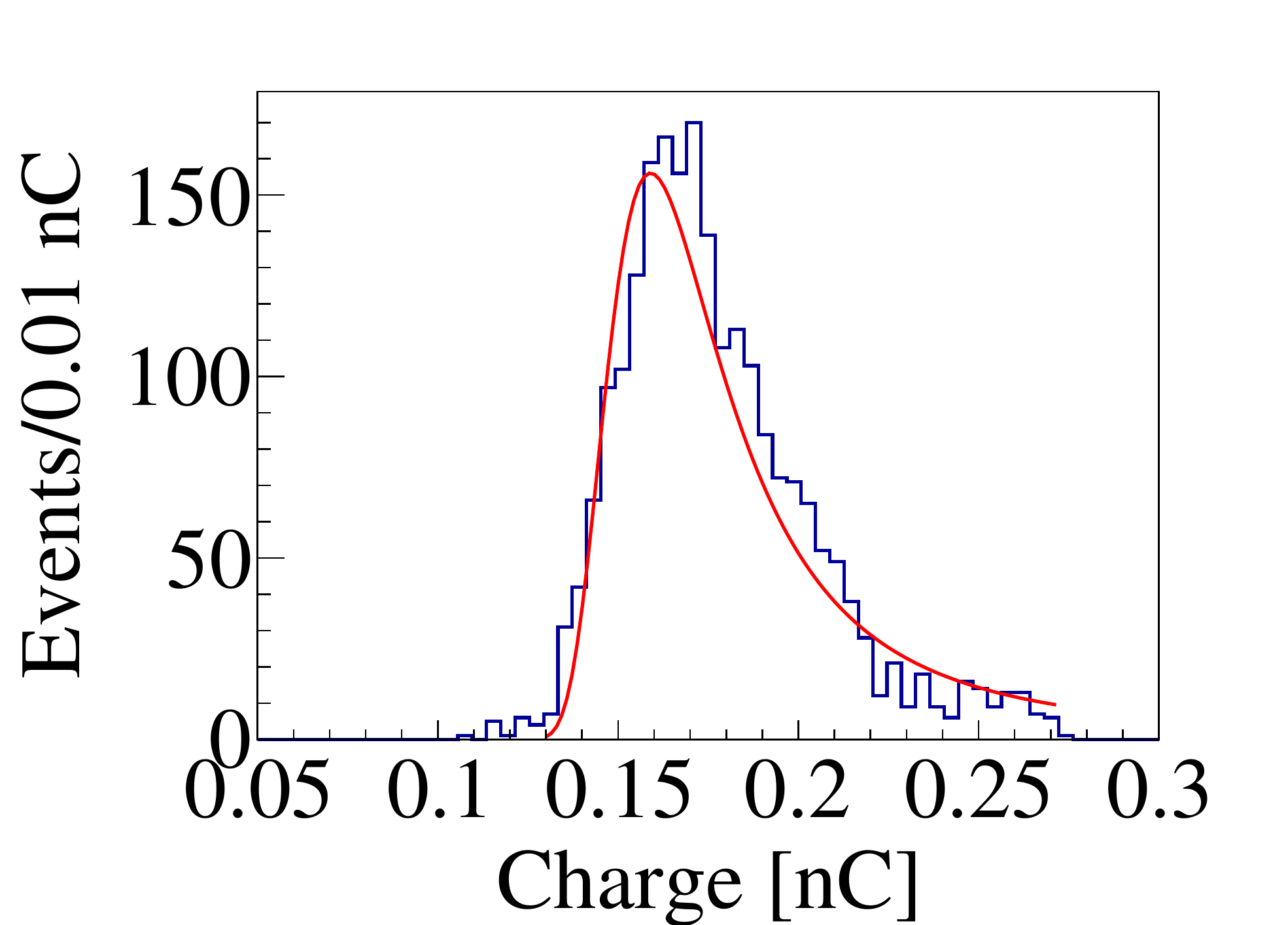}
\end{center}
\caption{Landau fittings of the charges in the four coincidence scintillators.
The gain of each channel was different.}
\label{Landau}
\end{figure}

\subsection{Average waveforms}
\begin{figure}[!h]
\begin{center}
\includegraphics[width=8cm]{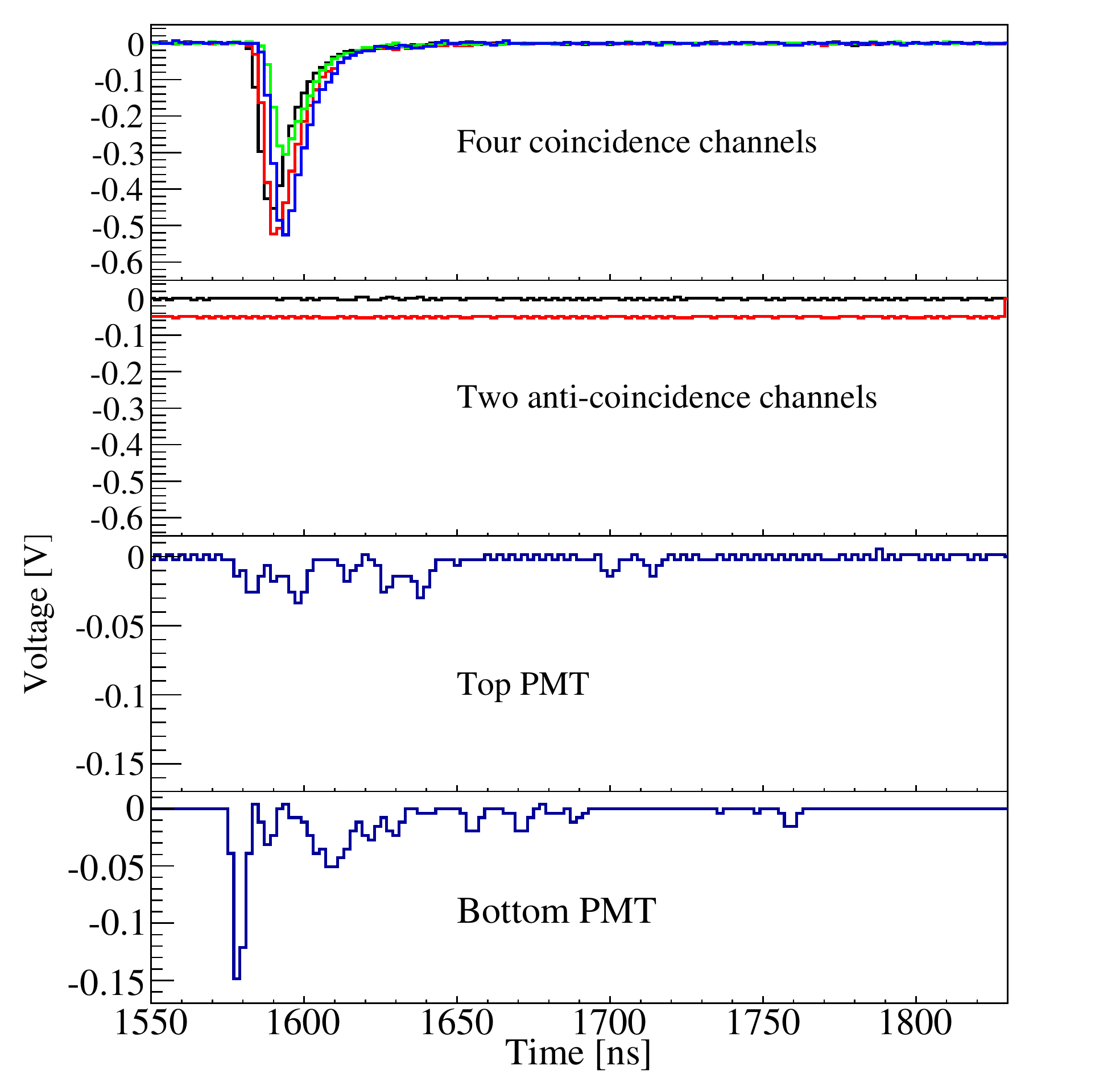}
\end{center}
\caption{Waveforms of the eight channels of a selected event. Waveforms of the two anti-coincidence scintillators were vertically shifted to avoid overlapping.
The bottom PMT observed both scintillation and Cherenkov lights and
the top PMT detected no obvious Cherenkov light.}
\label{8SignalsInOneEvent}
\end{figure}

\begin{figure}[!h]
\begin{center}
\includegraphics[width=8cm,clip]{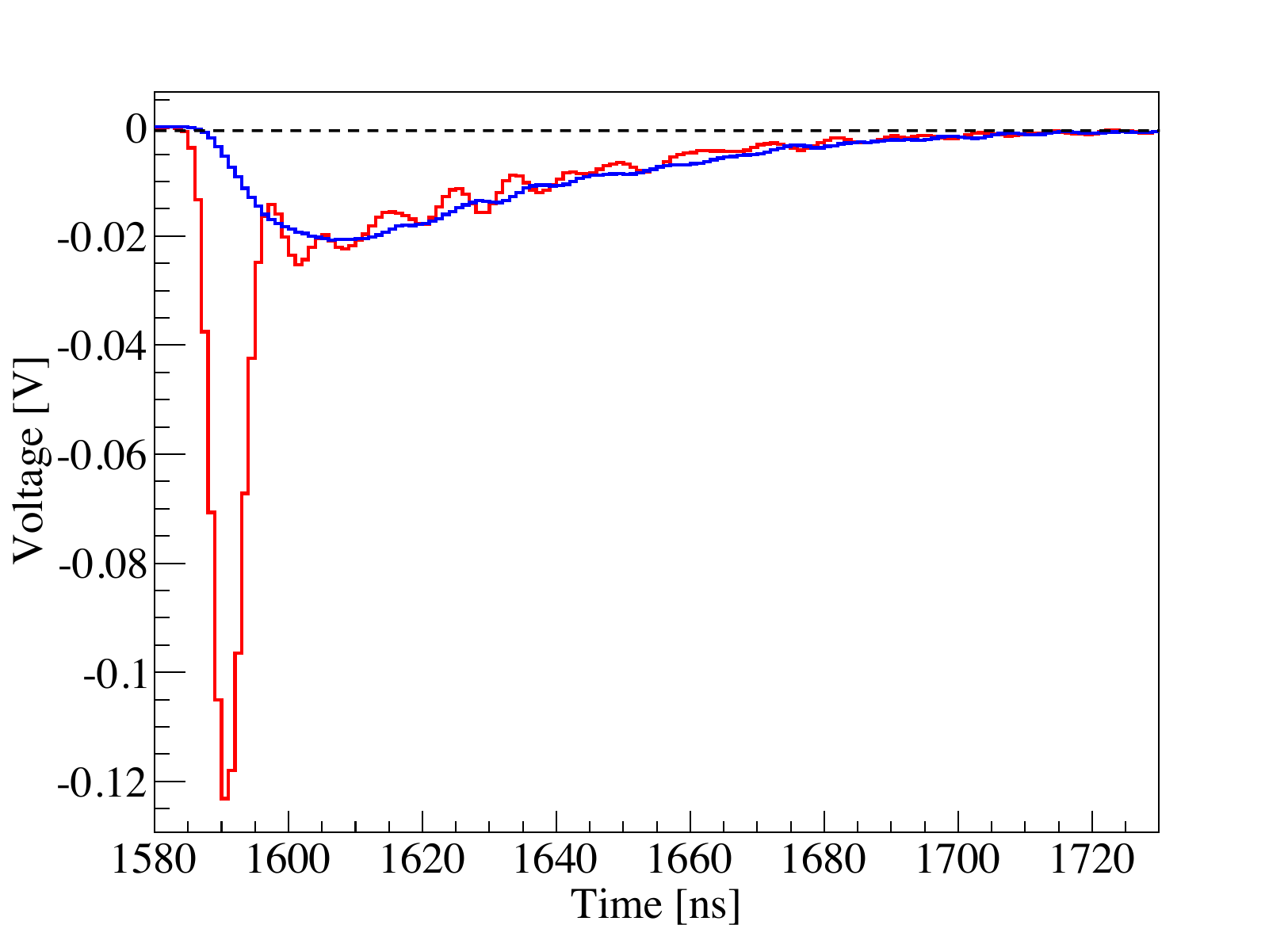}
\end{center}
\caption{Average waveforms of the top (blue) and bottom (red) PMTs.
The bottom PMT observed both scintillation and Cherenkov lights and
the top PMT detected no obvious Cherenkov light.
The extra oscillation structure, especially in the bottom channel, was consistent with the PMT's baseline ringing,
whose amplitude was less than 3 mV.}
\label{AverageAccumulateWave}
\end{figure}

After the signal selection, more than 2,000 candidates survived and the event rate was about 1.7/min. Figure~\ref{8SignalsInOneEvent} shows the waveforms of the eight channels of a candidate event, where extra activities
on the top and bottom PMT's were seen even 100 ns after the trigger channels.
The average waveform of the top and bottom PMT's are shown in Fig.~\ref{AverageAccumulateWave}.
A prompt Cherenkov peak can be found for the bottom PMT, however this peak was absent for the top PMT. A long symmetric tail of scintillation light were observed in both channels.
The extra oscillation structure, especially in the bottom channel, was consistent with the baseline ringing of the PMT, whose amplitude was less than 3 mV.

\section{Detector Simulation}
\label{sec:Simulation}

The whole apparatus was simulated to validate our understanding of the collected data and to estimate the efficiency of scintillation-light detection.

The detector geometry, the interaction of particles with materials, and the optical photons' propagation in the container were simulated by Geant4~\cite{g41, g42}.
Cosmic-ray muons, the quenching effect, time profile and light yield of scintillation, PMT response, etc.~were simulated by our customized program, and are explained below.

\subsection{Muon generator and event rate}
The simulation started with a cosmic-ray muon generator.
The energy and zenith angle of cosmic-ray muons was assumed to follow the modified Gaisser formula~\cite{Gaisser, GaisserM}
\begin{equation}
\label{eq:muondist2}
\begin{aligned}
\frac{\ud I^\star}{\ud E_\mu\ud\Omega}&=\frac{0.14(E_\mu^\star)^{-2.7}}{\textrm{cm}\cdot\textrm{s}\cdot\textrm{sr}\cdot\textrm{GeV}}\\
&\phantom{=}\times\left(\frac{1}{1+\frac{1.1E_\mu\cos\theta^\star}{115\textrm{GeV}}}+\frac{0.054}{1+\frac{1.1E_\mu\cos\theta^\star}{850\textrm{GeV}}}\right),
\end{aligned}
\end{equation}
where $I^\star$ is the muon flux, $E_\mu^\star$ and $\theta^\star$ are the corrected terms of muon kinetic energy and zenith angle, respectively.

Muons started on the top surface of the first coincidence scintillator and only 0.8\% of them could satisfy the quadruple-coincidence requirement.
The muon flux at sea level is about $1/(\rm{{cm}^{2}min})$~\cite{PDG} and the simulated event rate was 1.8/min, which was basically consistent with our measurement.

\subsection{Quenching effect}
Organic scintillators, like LAB, do not respond linearly to the ionization density.
Birks' Law~\cite{birks} is a semi-empirical formula to describe the quenching effect
\begin{equation}
dE_\text{vis}/dx=\frac{dE/dx}{1+k_BdE/dx},
\label{birk}
\end{equation}
where $dE_\text{vis}/dx$ is the visible energy loss density for scintillation light generation, $dE/dx$ is the actual energy loss density of a charged particle, and $k_B$ is the Birks' constant.
The Birks' constants for low energy electrons from~\cite{birksMea} were used in the simulation,
and $dE_\text{vis}/dx$ was decreased by $\sim2.8\%$ with respect to $dE/dx$ if $k_B$ was set to 0.015 cm/MeV.
The mean of the total visible energies in the container in the LAB was estimated to be $(64.2\pm1.8)\ \rm{MeV}$,
where the error was set to the range without quenching effect.

\subsection{Reflectivity}
The diffuse reflectivity of the inner surface of the container was scanned from 0\% to 5\% in the simulation,
The efficiencies estimated with 5\% reflectivity were used as central values.
It introduced less than 6\% uncertainty to the efficiency estimation of scintillation-light detection.

\subsection{LAB emission spectrum and light yield}
\begin{figure}[!h]
\includegraphics[width=8cm]{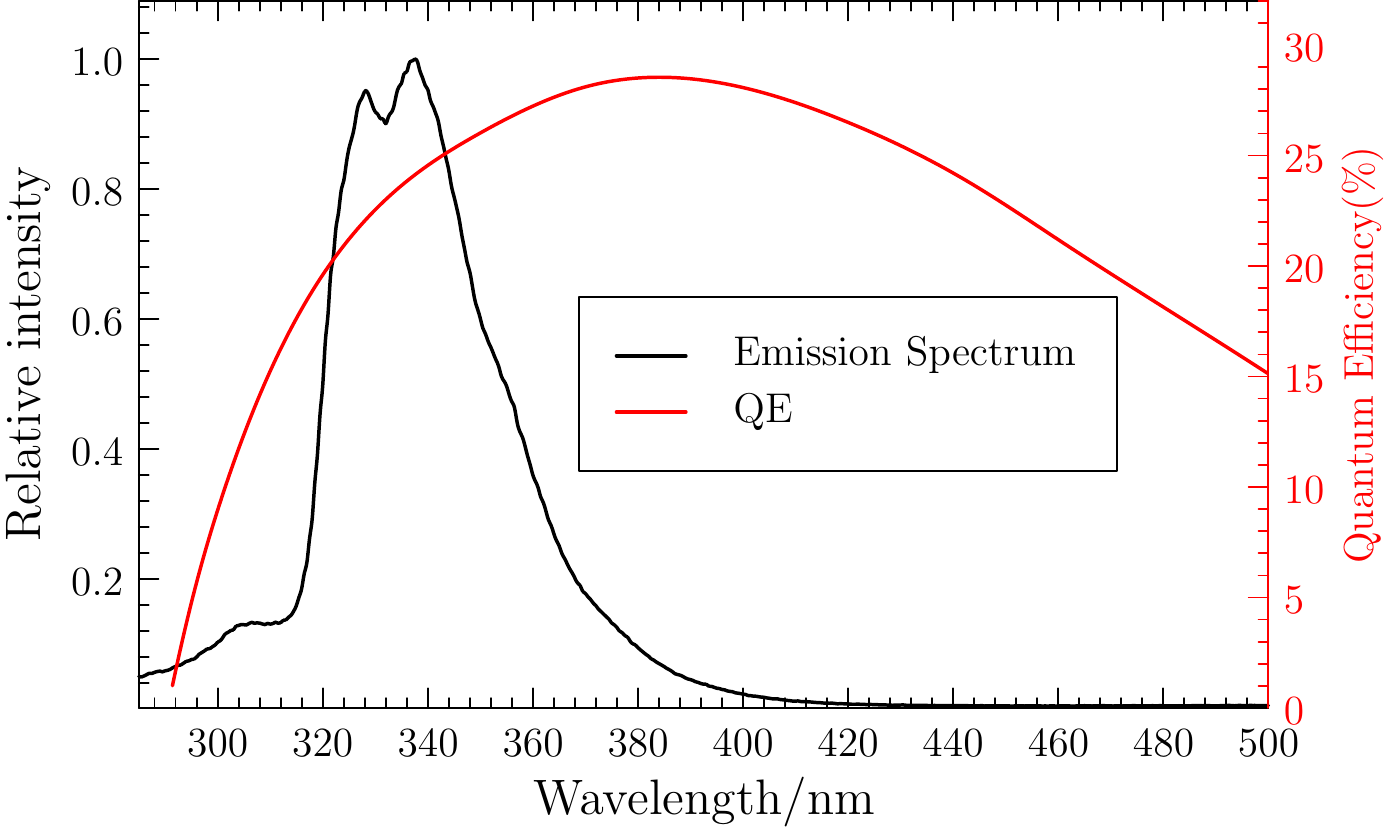}
\caption{The emission spectrum of linear alkyl benzene and the quantum efficiency of the PMT's in the related range of wavelength.}
\label{LAB_emission}
\end{figure}

The emission spectrum of linear alkyl benzene (10\% in cyclohexane) was measured using an RTI fluorescence
spectrometer excited at 280 nm with 1nm optical slit.
The spectrum is shown in Fig.~\ref{LAB_emission} and implemented accordingly in the simulation.
The scintillation light yield of LAB was set to our measurement reported later.

\subsection{PMT quantum and collection efficiency}
When optical photons reached the top and bottom PMT's photocathodes in our simulation, they were converted to photoelectrons
according to the product of the quantum efficiency, $\varepsilon_{QE}$, and collection efficiency, $\varepsilon_{CE}$, of the PMT.
The quantum efficiency as a function of wavelength was simulated according to the Hamamatsu data sheet~\cite{Hama},
which is over 10\% from 300 nm to 530 nm.
The collection efficiency is about 80\%~\cite{Hama}.
The total uncertainty for $\varepsilon_{QE}\times\varepsilon_{QE}$ was assumed to be 10\%.

\subsection{Detected wavelength spectrum}
\begin{figure}[!h]
\includegraphics[width=0.45\textwidth]{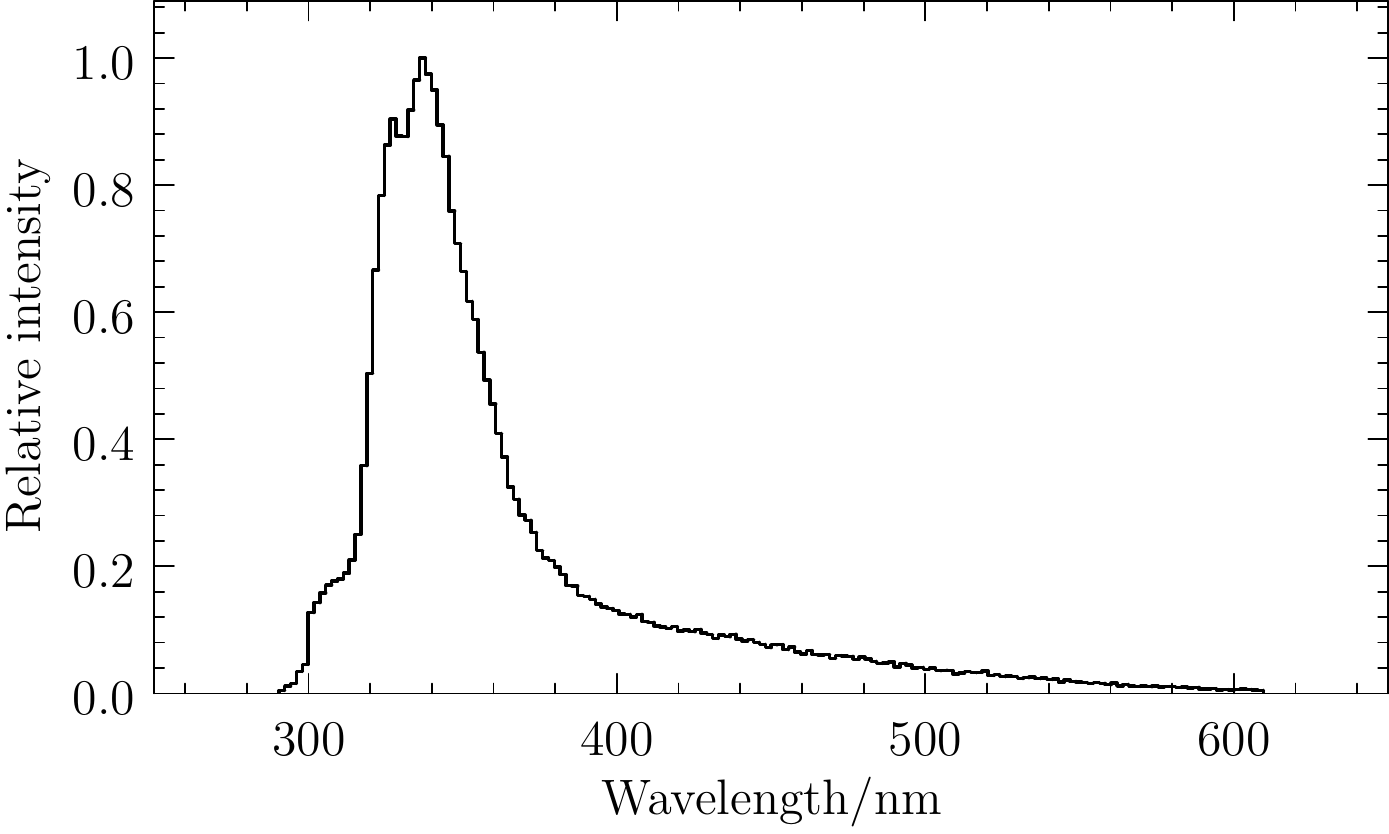}
\caption{Wavelength spectrum of all the detected photons in the simulation. The peak at 340 nm represented the scintillation photons, and the Cherenkov photons had a continuing spectra from 300\,nm to 600\,nm, where the light yield of the scintillation light were set according our measurement.}
\label{epd}
\end{figure}

The wavelength spectrum of all optical photons which finally were converted to photoelectrons
is shown in Fig.~\ref{epd}.

\subsection{Efficiency estimation}
The efficiency is defined as
\begin{equation}
\varepsilon=\frac{D}{N},
\label{effi}
\end{equation}
where $D$ is the number of detected scintillation or Cherenkov photoelectrons on a PMT and $N$ is
the total number of the corresponding photons.
The estimation for scintillation and Cherenkov lights on top and bottom PMT's were summarized in table~\ref{eff}.
The uncertainly came from the PMT quantum and collection efficiencies and the reflectivity of the inner surface of the container.

The number of detected Cherenkov photoelectrons of the bottom PMT in the simulation was $11.4\pm1.3$,
which is 30 times higher than that on the top PMT.
The conclusion is consistent with the purpose of the detector.
\begin{table}[!h]
\begin{tabular}{ccc}
\hline
               & Top PMT  & Bottom PMT\\
\hline
Cherenkov      & $(0.21\pm0.14)\times10^{-4}$   & $(6.32\pm0.70)\times10^{-4}$\\
Scintillation  & $(2.74\pm0.33)\times10^{-4}$   & $(2.73\pm0.33)\times10^{-4}$\\
\hline
\end{tabular}
\caption{Detection efficiencies of the Cherenkov and scintillation lights estimated by the simulation.}
\label{eff}
\end{table}

\section{Experiment results}
\label{sec:Results}

\subsection{Time profile of Scintillation light}
Given the detected waveforms in Fig.~\ref{AverageAccumulateWave},
the average pulse shape of the scintillation light is interpreted as
\begin{equation}
n(t)=\frac{\tau_r+\tau_d}{\tau_d^2}(1-\ue^{-t/\tau_r})\cdot\ue^{-t/\tau_d},\qquad t>0,
\label{sciresp}
\end{equation}
where $\tau_r$ is a rising time constant, $\tau_d$ is a decay time constant, and
$n(t)$ is normalized.

The top and bottom wavefroms were fitted with equation~\ref{sciresp} convoluted with a time response function with a Gaussian resolution plus a prompt Gaussian for the Cherenkov light with the same resolution. The formulas are
\begin{equation}
\begin{aligned}
f_T(t) &= A_{C,T}~\rm{gaus}(t_{\mu,T},\sigma) + A_{S,T}~n(t-t_0) \otimes \rm{gaus}(\sigma),\\
f_B(t) &= A_{C,B}~\rm{gaus}(t_{\mu,B},\sigma) + A_{S,B}~n(t-t_0) \otimes \rm{gaus}(\sigma),
\end{aligned}
\label{eq:up}
\end{equation}
where $A_C$ is the amplitude for the Cherenkov light, $t_{\mu}$ is the peak position of that,
$\sigma$ is the resolution of the time response function,
$A_S$ is the amplitude for the scintillation light, and $t_0$ is to account for the unknown starting time of the
scintillation emission, and subscript $T$ and $B$ are for top and bottom PMT's, respectively.

\begin{figure}[!h]
\includegraphics[width=0.45\textwidth]{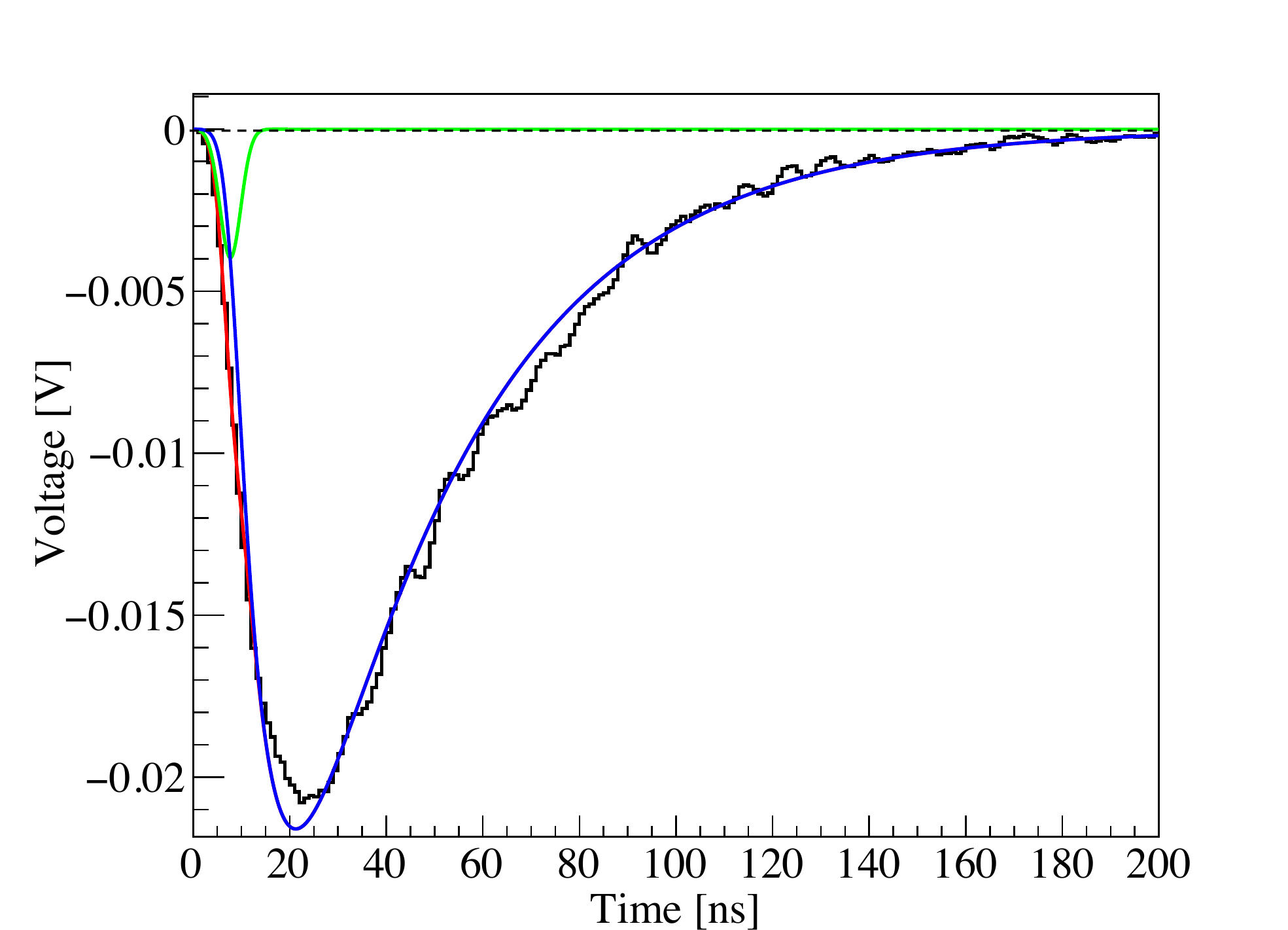}
\includegraphics[width=0.45\textwidth]{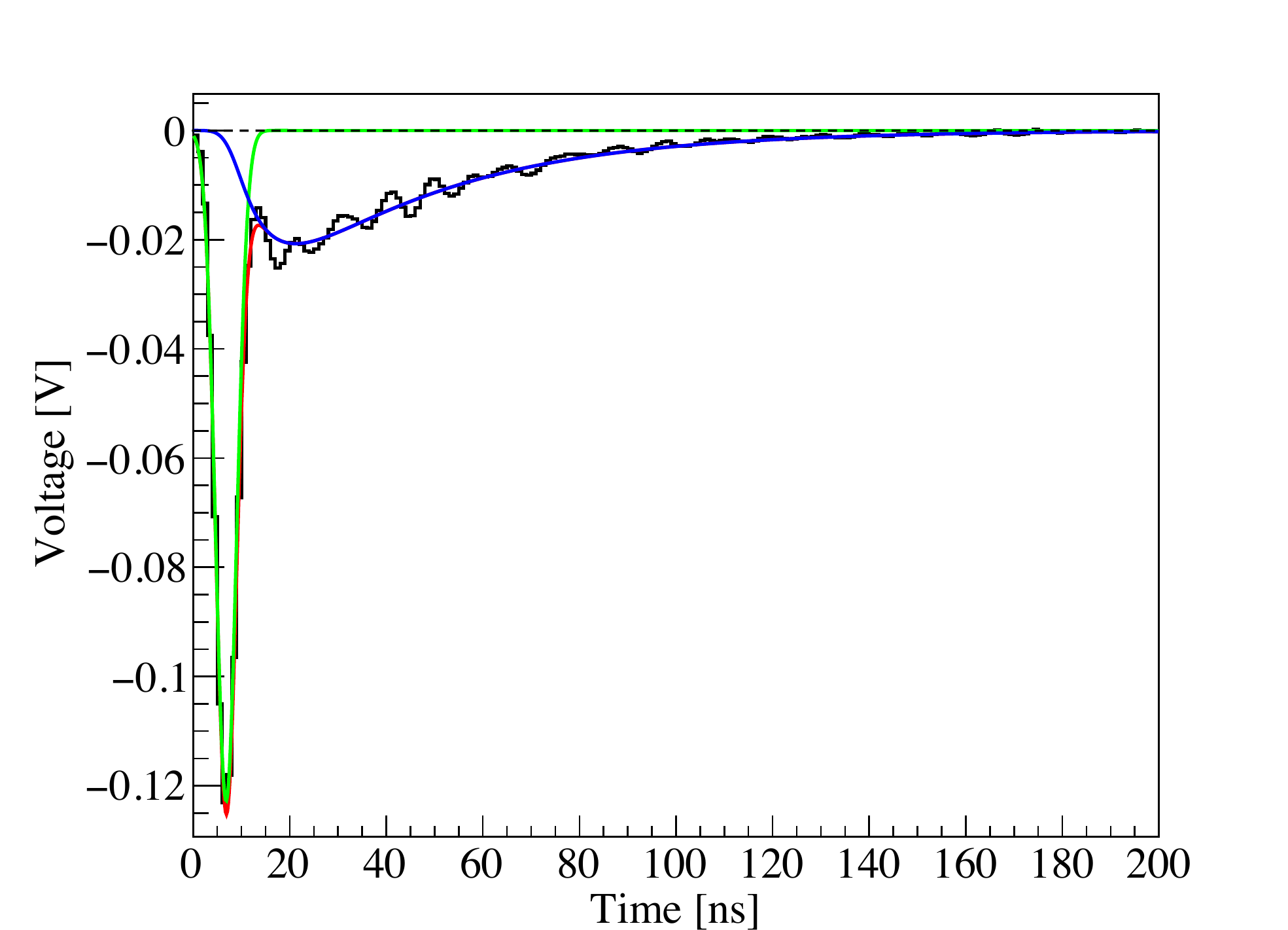}
\caption{Time profile of the scintillation and Cherenkov lights on the top (upper plot) and bottom (lower plot) PMT's and their fit results,
where the green is for the Cherenkov component, the blue for the scintillation component, and the red for their sum.}
\label{fitres}
\end{figure}

The top and bottom PMT's were fitted simultaneously, and the $t_{\mu,T}$ is constrained within 6-9 ns,
which is slightly later than the top one considering reflections.
An uncorrelated 3 mV uncertainty caused by the PMT ringing was conservatively introduced to each time bin in the fit.
As shown in Fig.~\ref{fitres} of the fitting results, the rising time and decay time observed are $\rt$ and $\dt$, respectively. The full width [-3$\sigma$, 3$\sigma$] of the Cherenkov light is 12 ns dominated by the time resolution of our PMT's.

\subsection{Scintillation light yield}
We calibrated the gain of the PMT's, and with the efficiency estimated by the simulation, we derived the scintillation light yield of the LAB.

Under the same high voltage as the data taking, dark noise events were collected. The gains of the top and bottom PMT's were estimated as $(8.2 \pm 0.1) \times 10^6$ and $(7.8 \pm 0.1) \times 10^6$, respectively, where the errors were statistical. The floating of the PMT gain is less than 2\%.

The number of photoelectrons on the top and bottom PMT's were extracted from the fits in the previous section and the gains measured. As shown in table~\ref{expPE},
the number of photoelectrons of the Cherenkov light for the bottom PMT is consistent with our simulation estimation in section~\ref{sec:Simulation}, and thus providing a validation of the simulation.

\begin{table}[htbp]
\begin{tabular}{ccc}
\hline
              & Top PMT  & Bottom PMT\\
\hline
Cherenkov     & $0.33\pm0.33$ & $10.7\pm0.4$\\
Scintillation & $17.6\pm0.6 $ & $17.7\pm0.6$\\
\hline
\end{tabular}
\caption{Measured photoelectrons of the scintillation and Cherenkov lights of the top and bottom PMT's. }
\label{expPE}
\end{table}

The scintillation light yield of the LAB was calculated with the following relation
\begin{equation}
L = \frac{D_{s,\text{exp}}}{\varepsilon_{s,\text{sim}} E_\text{vis}},
\end{equation}
where $D_{s,\text{exp}}$ is the number of measured photoelectrons of scintillation on the top PMT,
$\varepsilon_{s,\text{sim}}$ is the detection efficiency of scintillation light in table~\ref{eff},
and $E_\text{vis}$ is the total visible energy estimated by the simulation.
Taking the measurements and simulations of the top PMT,
the light yield was measured to be $\ly$.

\section{Conclusion and outlook}
\label{sec:Conclusion}
In this paper we observed a good separation of the scintillation and Cherenkov lights in an LAB sample.
The rising and decay times of the scintillation were measured to be $\rt$ and $\dt$, respectively,
and its light yield was measured to be $\ly$.

The long time constants of the LAB provide an opportunity to separate scintillation and Cherenkov lights as shown in Fig.~\ref{AverageAccumulateWave}. However the scintillation light yield is much lower than the liquid scintillators currently used for neutrino experiments, and it is very low for low energy solar neutrino experiments, where high energy resolution is expected.

According to the wavelength spectrum of our simulations, Fig.~\ref{epd}, wavelength shifters may be added to further increase the wavelength of the scintillation light emission and to enhance the scintillation light yield.
When some wavelength shifter added, the response time of LAB may become faster. This needs faster PMT's or new techniques like LAPPD~\cite{LAPPD2} to ensure a separation of Cherenkov and scintillation lights.

\section*{Acknowledgement}
We'd like to thank the support from
Key Laboratory of Particle \& Radiation Imaging (Tsinghua University), Ministry of Education,
Tsinghua University Initiative Scientific Research Program (No.~2012Z02161),
and Natural Science Foundation of China (No.~11235006 and No.~11475093),
and
portion of this work performed at Brookhaven
National Laboratory is supported in part by the United States
Department of Energy under contract DE-AC02-98CH10886.


\begin{thebibliography}{99}

\bibitem{snowmass} M.~Demarteau {\it et al.},  arXiv:1401.6116 (2014).
\bibitem{IMB}      C.~B.~Bratton {\it et al.}, Phys.Rev. D37 (1988) 3361.
\bibitem{SK}       S.~Fukuda {\it et al.}, Nuclear Instruments and Methods in Physics Research A 501 (2003) 418.
\bibitem{SNO}      J.~Boger {\it et al.}, Nuclear Instruments and Methods in Physics Research A 449 (2000) 172.
\bibitem{KamLAND}  K.~Eguchi {\it et al.}, Phys. Rev. Lett. 90 (2003) 021802.
\bibitem{Borexino} G.~Alimonti {\it et al.}, Nuclear Instruments and Methods in Physics Research A 600 (2009) 568.
\bibitem{LSND} C.~Athanassopoulos {\it et al.}, Nuclear Instruments and Methods in Physics Research A 388 (1997) 149.
\bibitem{DC} M.~Apollonio {\it et al.}, Eur. Phys. J. C 27 (2003) 331.
\bibitem{RENO} J.~K.~Ahn {\it et al.}, Phy. Rev. Lett. 108 (2012) 191802.
\bibitem{DYB} F.~P.~An {\it et al.}, Chin. Phys. C 37, (2013) 011001.

\bibitem{IEEE} D.~R.~Winn and D.~Raftery, IEEE Trans. Nucl. Sci. 32 (1985) 727.
\bibitem{Minfang} M.~Yeh {\it et al}., Nucl. Instr. Meth. A 51 (2011) 660.
\bibitem{Charac} L.~J.~Bignella {\it et al}., arXiv:1508.07029 (2015).
\bibitem{Damage} L.~J.~Bignella {\it et al}., arXiv:1508.07023 (2015).
\bibitem{Korea} S.~H.~So {\it et al}., Advances in High Energy Physics, 2014 (2014) 327184.
\bibitem{ASDC} J.~R.~Alonso {\it et al}., aiXiv:1409.5864 (2014).
\bibitem{THEIA} G.~D.~Orebi Gann {it et al}., arXiv:1504.02154 (2015).

\bibitem{BX1} G.~Bellini {it et al}., Phys. Rev. D 89 (2014) 112007.

\bibitem{PDG} K.~A.~Olive {\it et al}. (Particle Data Group), Chin. Phys. C 38 (2014) 090001.

\bibitem{Hama} Hamamatsu Photonics, see also the web link http://www.hamamatsu.com/.

\bibitem{g41} S.~Agostinelli {\it et al}., Nuclear Instruments and Methods in Physics Research A 506 (2003) 250.
\bibitem{g42} J.~Allison {\it et al}., IEEE Trans. Nucl. Sci. 53 (2006) 270.
\bibitem{Gaisser} T.~K.~Gaisser, Cosmic Rays and Particle Physics, Cambridge (1990).
\bibitem{GaisserM} D.~Chirkin, arXiv:hep-ph/040707 (2004).

\bibitem{birks} J.~B.~JBirks, The Theory and Practice of Scintillation Counting. London: Pergamon Press (1964).
\bibitem{birksMea} C.~Aberle {\it et al}., JINST 6, P11006 (2011).

\bibitem{LAPPD2} B.~Adams {\it et al}., Nuclear Instruments and Methods in Physics Research A 732 (2013) 392.

\end{thebibliography}
\end{document}